\documentclass[12pt,a4paper]{article}
\usepackage{mathtools}
\usepackage{mathrsfs}
\usepackage{amsmath}
\usepackage{tikz-feynman}
\usepackage[utf8x]{inputenc}
\usepackage{amssymb}
\usepackage{slashed}
\usepackage{bbm}
\usepackage{bm}
\usepackage{color}
\usepackage[a4paper,top=3cm,bottom=3cm,left=2cm,right=3cm,bindingoffset=5mm]{geometry}
\usepackage{booktabs} 
\usepackage{tikz}
\usepackage{cancel}
\usepackage[bottom]{footmisc}
\usepackage{cite}
\usepackage{float}
\usepackage{todonotes}
\usepackage{jheppub}

\usepackage{hyperref}

\hypersetup{
colorlinks=true,         
linkcolor=blue,          
citecolor=red,        
urlcolor=red            
}

\renewcommand{\[}{\begin{equation}\begin{aligned}}
\renewcommand{\]}{\end{aligned}\end{equation}}

\newcommand{\zb}{{\bar{z}}}
\newcommand{\wb}{{\bar{w}}}

\newcommand{\Ab}{{\bar{A}}}

\title{Celestial chiral algebras, colour-kinematics duality and integrability}
\author{Ricardo Monteiro}

\affiliation{Centre for Theoretical Physics, Department of Physics and Astronomy, \\
Queen Mary University of London, E1 4NS, United Kingdom}

\abstract{
We study celestial chiral algebras appearing in celestial holography, using the light-cone gauge formulation of self-dual Yang-Mills theory and self-dual gravity, and explore also a deformation of the latter. The recently discussed $w_{1+\infty}$ algebra in self-dual gravity arises from the soft expansion of an area-preserving diffeomorphism algebra, which plays the role of the kinematic algebra in the colour-kinematics duality and the double copy relation between the self-dual theories. A $W$ deformation of $w_{1+\infty}$ arises from a Moyal deformation of self-dual gravity. This theory is interpreted as a constrained chiral higher-spin gravity, where the field is a tower of higher-spin components fully constrained by the graviton component. In all these theories, the chiral structure of the operator-product expansion exhibits the colour-kinematics duality: the implicit `left algebra' is the self-dual kinematic algebra, while the `right algebra' provides the structure constants of the operator-product expansion, ensuring its associativity at tree level. In a scattering amplitudes version of the Ward conjecture, the left algebra ensures the classical integrability of this type of theories. In particular, it enforces the vanishing of the tree-level amplitudes via the double copy.
}

\begin{document}

\begin{flushright}
QMUL-PH-22-26
\end{flushright}

\maketitle

\section{Introduction}
\label{sec:intro}

In recent years, a new understanding of asymptotic symmetries in quantum field theory and quantum gravity for asymptotically flat spacetimes has been developed, starting with a connection to soft theorems obeyed by scattering amplitudes \cite{Strominger:2013jfa,He:2014laa,Strominger:2017zoo}. This led to a renewed effort to understand holography in asymptotically flat spacetimes, particularly in 4D, based on the interpretation of scattering amplitudes as 2D CFT correlation functions on the celestial sphere, as reviewed in \cite{Raclariu:2021zjz,Pasterski:2021rjz,Pasterski:2021raf,McLoughlin:2022ljp}. A recent discussion of how this picture of a 2D CFT  relates to an expected 3D holographic dual of 4D physics can be found in \cite{Donnay:2022aba}. 

The soft tower of asymptotic symmetries \cite{Donnay:2018neh,Adamo:2019ipt,Puhm:2019zbl,Guevara:2019ypd} can be organised in a particularly elegant manner in the self-dual --- i.e.~positive-helicity --- sectors of 4D Yang-Mills theory and gravity \cite{Guevara:2021abz}. In \cite{Strominger:2021lvk}, it was shown that the soft tower in self-dual gravity can be arranged into the $w_{1+\infty}$ algebra (more precisely, into a wedge subalgebra of $w_{1+\infty}$). Since this algebra is related to the behaviour of amplitudes in a collinear limit, and this behaviour is unaltered at loop level in the self-dual sector, the conclusion extends to quantum self-dual gravity \cite{Ball:2021tmb}. A celestial realisation of the $w_{1+\infty}$ algebra was presented in \cite{Adamo:2021lrv}, based on the twistor space encoding of self-dual gravity. See also \cite{Himwich:2021dau,Jiang:2021csc,Mago:2021wje,Freidel:2021ytz,Costello:2022wso,Costello:2022upu,Ren:2022sws,Compere:2022lzx} for related work.

In this paper, we show how the formulations of the self-dual theories in light-cone gauge -- arguably their simplest spacetime formulations --- illuminate the appearance of the $w_{1+\infty}$ algebra, as well as that of a natural deformation of this algebra. The latter reveals a connection to the chiral higher-spin theories developed in \cite{Metsaev:1991mt,Metsaev:1991nb,Ponomarev:2016lrm}.

Our aproach reveals a strong connection between celestial operator-product expansions (OPEs) and the topic of the double copy, which relates most notably Yang-Mills theory and gravity, but extends to a wide range of theories. The double copy has received much attention over the past decade or so, mainly due to its applications, from the structure of superstring perturbation theory to the calculation of observables in gravitational wave physics; see e.g.~\cite{Bern:2019prr,Bern:2022wqg,Kosower:2022yvp,Adamo:2022dcm} for recent reviews, and \cite{Pasterski:2020pdk,Casali:2020vuy,Casali:2020uvr,Campiglia:2021srh,Godazgar:2021iae,Adamo:2021dfg,Gonzo:2022tjm,Huang:2019cja,Alawadhi:2019urr,Banerjee:2019saj} for works motivated by celestial holography. The self-dual theories provide the simplest known setting for the double copy, by manifestly obeying the symmetry known as colour-kinematic duality \cite{Monteiro:2011pc,Boels:2013bi}. As we will see, this is closely related to the chirality of the celestial OPEs, and to the (classical) integrability of the chiral sector. The associativity of the OPEs, which implies the Jacobi relation for the structure constants, is also related to the appearance of Jacobi relations in the colour-kinematics duality, as we will see in examples.

The paper is organised follows. We start by giving a brief introduction to the spacetime formulations of self-dual Yang-Mills theory and self-dual gravity in light-cone gauge in section~\ref{sec:SD}. In the following sections, we discuss the relation between celestial chiral OPEs and the double copy structure of the self-dual theories. In section~\ref{sec:OPE},
we show how the first copy of the self-dual kinematic algebra underlying the colour-kinematics duality is related to the chirality of the OPEs, while the second copy (present in the gravity case) is related via a soft expansion to the $w_{1+\infty}$ algebra. We also discuss in that section a Moyal deformation of self-dual gravity, which leads to a deformation of the $w_{1+\infty}$ algebra, and its connection to higher-spin theories. In section~\ref{sec:int}, we describe the role of the first copy of the self-dual kinematic algebra in the classical integrability of the chiral theories, from the point of view of vanishing tree-level scattering amplitudes. Finally, we briefly discuss the results in section~\ref{sec:discussion}, with an eye to assessing the role of the $w_{1+\infty}$ algebra beyond the self-dual sector.

{\it Note:} Shortly after the first arXiv version of this paper, two papers appeared which have some overlap \cite{Bu:2022iak,Guevara:2022qnm}.

\section{Self-dual theories in light-cone gauge}
\label{sec:SD}

Self-dual Yang-Mills theory (SDYM) and self-dual gravity (SDG) admit well-known formulations in light-cone gauge that, while breaking manifest Lorentz invariance, are based only on the physical degrees of freedom. Let us work with light-cone coordinates $(u,v,w,\wb)$, such that the wave operator is $\,\square=\partial_u\partial_v-\partial_w\partial_\wb\,$. The action for SDYM can be written as \cite{Chalmers:1996rq} (see also \cite{PARK1990287,Cangemi:1996rx,Bardeen:1995gk})
\[
\label{eq:SDYM}
S_\text{SDYM}(\Psi,\bar\Psi) = \int d^4 x \;\; \text{tr}\; \bar\Psi \big(\square \Psi + i[\partial_u \Psi,\partial_w \Psi]\big) \,,
\]
where $\Psi$ and $\bar\Psi$ correspond to the positive and negative helicity degrees of freedom. The role of $\bar\Psi$ in the action is that of a Lagrange multiplier. The analogous action for SDG is \cite{Siegel:1992wd}
\[
\label{eq:SDG}
S_\text{SDG}(\phi,\bar\phi) = \int d^4 x \;\; \bar\phi \big(\square \phi + \{\partial_u \phi,\partial_w \phi\}\big)\,,
\] 
which makes use of a Poisson bracket on the null plane $(u,w)$,
\[
\label{eq:Poisson}
\{f,g\}:=\partial_u f \,\partial_w g -\partial_w f\, \partial_u g\,.
\]
The Lagrange multiplier $\bar\phi$ enforces Pleba\'nski's second heavenly equation \cite{Plebanski:1975wn}.

A comment for the reader unfamiliar with these actions: the interaction terms appear to have too many derivatives because, to tidy up the expressions, the helicity fields are chosen to be related to the relevant components of the gauge field and the graviton as
\[
\label{eq:rescale}
(A ,\Ab)= (\partial_u \Psi,\, \partial_u^{-1} \bar\Psi) \qquad \text{and} \qquad (h ,\bar h)= (\partial_u^2 \phi,\, \partial_u^{-2} \bar\phi)\,.
\]

One of the obvious similarities between the two actions is the analogy between the colour Lie bracket in \eqref{eq:SDYM} and the Poisson bracket in \eqref{eq:SDG}. This similarity becomes even more striking when we consider the momentum-space Feynman rules:
\begin{itemize}
\item Propagator $(+-)$:\, $\displaystyle \frac{i}{k^2}\,\delta^{ab}$.
\item Cubic vertex $(++-)$:
\[ 
\label{eq:3ptvert}
V_\text{SDYM} = X(k_1,k_2) \,f^{a_1a_2a_3}\,, \qquad V_\text{SDG} = X(k_1,k_2)^2 \,,
\]
 where\footnote{To actually compute scattering amplitudes, it is often convenient to keep track of the choice of light-cone using the spinor-helicity formalism. Following the notation of \cite{Boels:2013bi}, we can write $X(k_1,k_2)=\langle \eta | k_1k_2 | \eta\rangle$, where $|\eta\rangle [\eta|$ is the null vector defining the light-cone gauge. Moreover, a rescaling factor is required for each  $\pm$ helicity external gluon or graviton in the amplitude, equal to $\langle \eta i \rangle^{\mp 2}$ or $\langle \eta i \rangle^{\mp 4}$, respectively. These rules reduce to the ones in the main text if the spinor choices to be seen in \eqref{eq:kin} are followed, together with $|\eta\rangle=(0,1)$, so that we have $\langle \eta i \rangle^2=1$.\label{footnote}}
\[
X(k_1,k_2):=k_{1w}k_{2u}-k_{1u}k_{2w}=-X(k_2,k_1) \,.
\]
\end{itemize}
These kinematic objects are the structure constants of the Lie algebra
\[
\label{eq:LieXp}
\{e^{ik_1\cdot x},e^{ik_2\cdot x}\} = X(k_1,k_2) \,e^{i{(k_1+k_2)}\cdot x}\,.
\]
We can also write this algebra in terms of the Hamiltonian vector fields
\[
L_{k}:=\{e^{i\,k\cdot x},\cdot\}=i \,e^{i\,k\cdot x}(k_u\partial_w-k_w\partial_u)= e^{i\,k\cdot x}\,k_u\,\epsilon^+\!\cdot\partial
\]
as
\[
\label{eq:LieX}
[L_{k_1},L_{k_2}]= X(k_1,k_2)\,L_{k_1+k_2}\,.
\]
These vector fields generate area-preserving (i.e.~unit-Jacobian) diffeomorphisms in the $(u,w)$ plane, and $\epsilon^+(k)$ denotes a positive-helicity polarisation vector, defined whether or not $k$ is on-shell.

Notice that each of the 3-point vertices in \eqref{eq:3ptvert} has two sets of structure constants: the kinematic structure constant times the colour structure constant in SDYM, and two copies of the kinematic structure constant in SDG. As described in \cite{Monteiro:2011pc,Boels:2013bi}, this feature of the self-dual theories is a remarkably simple example of the BCJ colour-kinematics duality in Yang-Mills theory and the associated double copy to gravity \cite{Bern:2008qj,Bern:2019prr}. In general, the kinematic algebra underlying the colour-kinematics duality is much more sophisticated than its restriction to the self-dual sector, and does not admit a simple 3-point-vertex-type interpretation; see e.g.~\cite{Cheung:2021zvb,Brandhuber:2021bsf,Chen:2022nei,Brandhuber:2022enp} for recent studies.

SDYM and SDG are well-defined sectors of the complete 4D Yang-Mills theory and general relativity, obtained by restricting the interactions to the $(++-)$ vertex. The usual actions for the full theories have a $(++-)$ vertex, a $(+--)$ vertex, and higher-point vertices of the type $(++\cdots--)$. In the standard Yang-Mills action, the only higher-point vertex is the 4-point vertex $(++--)$. At tree level, the self-dual sector corresponds to helicity configurations where all external particles but one have positive helicity, in a convention where all external particles are defined as incoming (the momentum in the external legs goes into a Feynman diagram). The associated tree-level scattering amplitudes vanish, which is a sign of the classical integrability of the self-dual theories \cite{Bardeen:1995gk}.\footnote{The exception is the 3-pt amplitude, which in (1,3) spacetime signature has support only in complexified kinematics, but plays an important role in on-shell methods.\label{footnote3pts}} The idea is that no amplitude can be consistent with the infinite tower of symmetries underlying integrability. This is broken by quantum effects, which give rise to non-vanishing one-loop amplitudes, though the quantum self-dual theories are still very simple. It is not possible to draw a Feynman diagram with more than one loop using the Feynman rules above, which means that the self-dual theories are one-loop exact. Moreover, one-loop diagrams require all external particles to have positive helicity. This is the self-dual sector of the full theories at loop level. The one-loop all-positive-helicity amplitudes turn out to be rational functions of the external kinematic data \cite{Bern:1994zx,Bern:1998sv}. So the self-dual theories may be considered to be the simplest 4D theories with non-trivial amplitudes. It is important to notice, though, that self-dual Yang-Mills fields and metrics are complex in (1,3) spacetime signature; the theories are clearly not parity invariant. Another common point of view is to consider real self-dual fields in (2,2) signature. 

The structure of the chiral higher-spin theories studied in \cite{Metsaev:1991mt,Metsaev:1991nb,Ponomarev:2016lrm,Ponomarev:2017nrr}, which we aim to explore in future work, is closely related to that of SDYM and SDG. Similarly to these, the formulation in light-cone gauge is particularly simple, but they also admit a covariant description \cite{Krasnov:2021nsq,Tran:2021ukl,Skvortsov:2022syz,Sharapov:2022faa}.

\section{Celestial chiral OPEs}
\label{sec:OPE}

The main idea of celestial holography is that 4D scattering amplitudes should be thought of as correlation functions in a 2D conformal field theory living on the celestial sphere \cite{Cheung:2016iub,Pasterski:2016qvg,Pasterski:2017kqt,Pasterski:2017ylz}. The OPE of two operators in this celestial theory is determined by the collinear limit of scattering amplitudes \cite{Fan:2019emx,Pate:2019lpp}, as suggested by the figure.
\begin{center}
\includegraphics[width=2.5cm]{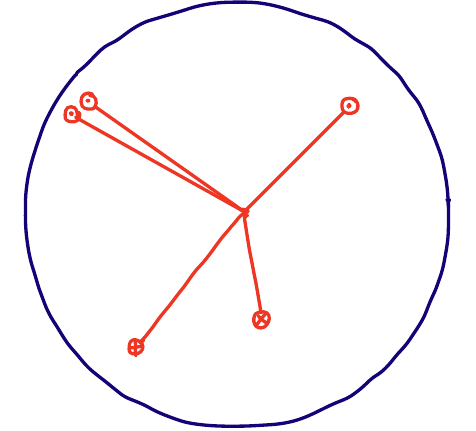}
\end{center}
While a basis of boost eigenstates is more appropriate to reveal the celestial conformal properties of scattering amplitudes, we will use in this paper momentum eigenstates. This choice connects more directly to common concepts in the amplitudes literature, which we want to make use of.

We parametrise the momenta of on-shell gluons or gravitons as
\[
\label{eq:kin}
k_{A\dot A}=\lambda_A\tilde\lambda_{\dot A}\,,
\quad \lambda_A=(1,z) \,, \quad \tilde\lambda_{\dot A}=(k_u,k_w) \,, \quad z=\frac{k_{\bar w}}{k_u}=\frac{k_v}{k_w}\,,
\]
where $z$ is a holomorphic coordinate on the celestial sphere. Notice that we used the scaling freedom $(\lambda_A\mapsto \tau \lambda_A\,,\, \tilde\lambda_{\dot A}\mapsto \tau^{-1} \tilde\lambda_{\dot A})$ to treat the two spinors differently. In particular, we have $\tilde\lambda_{\dot A}=k_u(1,\zb)$. With this choice, for on-shell momenta $k_1$ and $k_2$, we obtain
\[
X(k_1,k_2) = \epsilon^{\dot A \dot B}\,\tilde\lambda_{1\,\dot A}\,\tilde\lambda_{2\,\dot B}=:{ [12]} \,,
\]
and
\[
s_{12} :=(k_1+k_2)^2= (z_1-z_2)\, { [12]} \,.
\]

We now consider a `holomorphic collinear limit' by taking $z_1\to z_2$.\footnote{It is helpful to consider an analytic continuation in this context, allowing $z$ and $\zb$ to be independent. In fact, unlike (1,3) spacetime signature, (2,2) signature admits real independent $z$ and $\zb$, and also admits real self-dual gauge fields and metrics, making it a common setting to study self-dual theories. The counterpart of the celestial sphere is a celestial torus.} In this limit, the scattering amplitude is dominated by the pole $1/s_{12}$, whose residue is generically a sum (over on-shell states exchanged) of products of two subamplitudes: a 3-point subamplitude involving particles 1 and 2, and a subamplitude involving the remaining external particles. In both Yang-Mills theory and general relativity, if we take particles 1 and 2 to have positive helicity, then the 3-point subamplitude is associated to the vertex $(++-)$. The translation into celestial holography determines the OPEs of operators associated to positive-helicity scattering states.\footnote{The reason why the positive-helicity OPEs in the main text correspond to the $(++-)$ vertex is that the operators on the right-hand side of the OPE are `outgoing' with respect to the 3-pt vertex and have, therefore, negative helicity when see as incoming, which is our convention for the vertex.} For positive-helicity gluons in Yang-Mills theory, we have\footnote{One may ask how to think about the operators in these OPEs. We may pragmatically follow the approach of \cite{Fan:2019emx}: if it is true that correlation functions of operators in a two-dimensional celestial CFT reproduce 4D scattering amplitudes, then the OPEs must take this form dictated by the collinear limit of the amplitudes. Alternatively, one may be able to construct explicit representations for the operators as in the twistor space approach of \cite{Adamo:2021lrv}.}
\[
\label{eq:OPESDYM}
{\mathcal O}^{a_1}_+(k_1)\,\, {\mathcal O}^{a_2}_+(k_2) \,\sim\, \frac{{ X(k_1,k_2)} \,{ f^{a_1a_2a_3}}}{s_{12}} \,\,{\mathcal O}^{a_3}_+ (k_1+k_2)\,=\, \frac{{ f^{a_1a_2a_3}}}{z_1-z_2}\,\, {\mathcal O}^{a_3}_+(k_1+k_2)\,,
\]
while for positive-helicity gravitons in general relativity, we have
\[
{\mathcal O}_+(k_1)\,\, {\mathcal O}_+(k_2) \,\sim\, \frac{{ X(k_1,k_2)^2} }{s_{12}} \,\,{\mathcal O}_+ (k_1+k_2)\,=\, \frac{{ X(k_1,k_2)}}{z_1-z_2}\,\, {\mathcal O}_+(k_1+k_2)\,.
\]
We normalised the operators in analogy with the redefinition \eqref{eq:rescale}, which makes the expressions tidier. The $(++-)$ vertex appears in the self-dual sector \eqref{eq:3ptvert}, which is the reason why we are able to express these particular OPEs of the full theories in terms of the vertex of the self-dual theories. So, while the chiral OPEs above are embedded in the full theories (which have additional OPEs), considering the self-dual sector suffices to study the properties of these particular OPEs.\footnote{Notice that the $(++-)$ vertex also gives rise to a chiral OPE for mixed helicities in the self-dual theories, e.g.~${\mathcal O}_{+}(k_1)\, {\mathcal O}_{-}(k_2) \,\sim\,  \frac{ X(k_1,k_2)}{z_1-z_2}\,\, {\mathcal O}_{-}(k_1+k_2)$. In the full theories, however, the mixed-helicity OPE is not chiral, because it contains also an anti-chiral part, i.e.~a contribution from the $(--+)$ vertex.\label{footnote:chiral}}

It is important to notice that the chiral OPEs above share the structure of the colour-kinematics duality in the $(++-)$ vertex seen in \eqref{eq:3ptvert}. The chirality of the OPE, i.e.~the fact that we have poles $1/(z_1-z_2)$, is a consequence of the `left structure constant' $X(k_1,k_2)$ appearing in both Yang-Mills theory and gravity. The `right structure constant', namely $f^{a_1a_2a_3}$ for Yang-Mills theory and $X(k_1,k_2)$ for gravity, is identified with the structure constant of the OPE. Since the `right structure constants' obviously satisfy Jacobi identities, the OPE is associative at tree level.

\subsection{From the kinematic algebra to $w_{1+\infty}$ via soft expansion}

Following the preceding discussion, we note that
\[
\label{eq:OPEgrav}
{\mathcal O}_+(k_1)\,\, {\mathcal O}_+(k_2) \,\sim\, \frac{{ X(k_1,k_2)}}{z_1-z_2}\,\, {\mathcal O}_+(k_1+k_2)
\]
is a celestial chiral algebra version of the spacetime Lie algebra \eqref{eq:LieXp}, which we also expressed as \eqref{eq:LieX} in terms of the commutator of vector fields $L_k=\{e^{ik\cdot x},\,\cdot\,\}$.

In the following, we consider the soft expansion of \eqref{eq:OPEgrav}, first looking at the analogous soft expansion of \eqref{eq:LieXp}. In our choice of kinematics \eqref{eq:kin}, the soft limit $k\to0$ corresponds to $\tilde\lambda=(k_u,k_w)\to (0,0)$; at fixed $z$, this requires $(k_v,k_\wb) \to (0,0)$ with fixed $k_\wb/k_u=k_v/k_w=z$. We have, therefore, the expansion
\[
\label{eq:softexp}
e^{i\,k\cdot x}= \sum_{a,b=0}^\infty \frac{(ik_u)^a}{a!}\,\frac{(ik_w)^b}{b!}\;{ {\mathfrak e}_{a,b}}\,,
\quad\;\; {\mathfrak e}_{a,b} = (u+z\wb)^a(w+zv)^b \,.
\]
The counterpart of the kinematic algebra \eqref{eq:LieXp} in terms of the `soft mode generators' ${\mathfrak e}_{a,b}$ is
\[
\{ {\mathfrak e}_{a,b}\,,\, {\mathfrak e}_{c,d} \} = { (ad-bc)} \,{ {\mathfrak e}_{a+c-1,b+d-1}}\,.
\]
Alternatively, we can express this algebra in terms of the commutator of vector fields $\{{\mathfrak e}_{a,b},\,\cdot\,\}$.
This is precisely the wedge subalgebra of the Lie algebra $w_{1+\infty}$, where the `wedge' refers to the restriction $a,b\geq0$. We can also rewrite this exactly as in ref.~\cite{Strominger:2021lvk}:
\[
[\, w^p_m\,,\, w^q_n \,] = \big( m(q-1)-n(p-1)\big) \,{ w^{p+q-2}_{m+n}}\,,
\]
using \,$w^p_m=\{\frac1{2}\,{\mathfrak e}_{p-1+m\,,\,p-1-m}\,,\,\cdot\,\}$\,. The wedge condition is now $|m|\leq p-1$. 

From the correspondence between \eqref{eq:LieXp} and \eqref{eq:OPEgrav}, we can read off that the soft expansion at fixed $z$,
\[
{\mathcal O}_+(k)=\sum_{a,b=0}^\infty \frac{(ik_u)^a}{a!}\,\frac{(ik_w)^b}{b!}\;{ \varpi_{a,b}}(z)\,,
\]
leads to
\[
\label{eq:OPEsoftgrav}
\varpi_{a,b}(z_1)\, \varpi_{c,d}(z_2) \,\sim\, \frac{{ ad-bc}}{z_1-z_2}\,\,\varpi_{a+c-1,b+d-1}(z_2)\,. 
\]

This is a particularly simple derivation of the appearance of the $w_{1+\infty}$ algebra in SDG. We do not claim that this is an essentially new derivation: the relation between $w_{1+\infty}$ and area-preserving diffeomorphisms has a long history \cite{Bakas:1989xu,Hoppe:1988gk}, and the connection to SDG has been explored before, e.g.~in the context of twistor theory \cite{Boyer:1985aj,Park:1989fz,Park:1989vq,MasonN:1990,Dunajski:2000iq}, as recently reviewed  in  \cite{Adamo:2021lrv}. In that reference, an explicit celestial realisation of $w_{1+\infty}$ based on twistor space was presented. Our spacetime perspective is helpful for some purposes, such as the connection to the colour-kinematics duality as typically expressed in momentum space, which we will further elaborate on.

We expect that the role of the $w_{1+\infty}$ algebra in the full gravity theory is a counterpart of the role of the kinematic algebra associated to the $(++-)$ vertex: it rules the self-dual sector of the theory, and also some kinematic limits of the full theory that highlight this vertex as in the collinear limit considered above, but the structure of the full theory is much more intricate. In fact, we believe that an indication of the sophistication of the full theory is the difficulty in understanding the general kinematic algebra underlying the colour-kinematics duality, which generalises \eqref{eq:LieX} beyond the self-dual sector. We will make a few more comments on this point in the Discussion section.

\subsection{Moyal-deformed self-dual gravity, $W$ and chiral higher spins}

A deformation of SDG by higher-derivative interactions that preserves its classical integrability was introduced in \cite{STRACHAN199263}; see also \cite{Garcia-Compean:2003nix}. The double copy structure of the deformed theory was described recently in \cite{Chacon:2020fmr}. It appears that this deformation has received little attention, so we will also discuss here its physical interpretation, and find a connection to chiral higher-spin theories.

The deformation is based on substituting the Poisson bracket \eqref{eq:Poisson},
\[
\label{eq:PoissonP}
\{f,g\}=f \buildrel{\leftrightarrow} \over {P} g \,, \qquad 
\buildrel{\leftrightarrow} \over {P} \; = \;
\buildrel{\leftarrow} \over{\partial}_u
\buildrel{\rightarrow} \over{\partial}_w-
\buildrel{\leftarrow} \over{\partial}_w
\buildrel{\rightarrow} \over{\partial}_u\,,
\]
by the Moyal bracket
\[
\label{eq:Moyal}
\{f,g\}^M:=\frac{1}{2\alpha}(f\star g-g\star f) =\frac{1}{\alpha}\, f \sinh(\alpha \! \buildrel{\leftrightarrow} \over {P})\, g \,,
\]
which is based on the Moyal star product with deformation parameter $\alpha$,
\[
\label{eq:starprod}
f\star g:= f\exp\left(\alpha \! \buildrel{\leftrightarrow} \over {P}\right) g\,.
\]
We have
\[
f\star g \,\stackrel{\alpha\to0}{\longrightarrow}\, fg \qquad \text{and} \qquad 
\{f,g\}^M \,\stackrel{\alpha\to0}{\longrightarrow}\, \{f,g\}\,.
\]
In full detail,
\[
\label{eq:Moyalexpl}
\{f,g\}^{M}=\sum_{s=0}^{\infty}\frac{\alpha^{2s}}{(2s+1)!}\sum _{j=0}^{2s+1}(-1)^{j} \binom{2s+1}{j}(\partial_{u}^{2s+1-j}\partial_{w}^{j}f)(\partial_{w}^{2s+1-j}\partial_{u}^{j}\,g).
\]
Since the star product is associative, the Moyal bracket satisfies the Jacobi relation, and therefore defines a Lie algebra. In fact, the Moyal bracket is the most general bracket one can write down for a Lie algebra of functions of two variables \cite{Fletcher:1990ib}. In our case, those two variables are $u$ and $w$, since $v$ and $\wb$ are passive coordinates with respect to this algebraic structure.

The Moyal deformation of SDG has action
\[
\label{eq:MSDG}
S_\text{Moyal-SDG}(\phi,\bar\phi) = \int d^4 x \;\; \bar\phi \big(\square \phi + \{\partial_u \phi,\partial_w \phi\}^M\big)\,.
\] 
The 3-pt vertex is now
\[
\label{eq:3ptvertMSDG}
V_\text{Moyal-SDG} = X(k_1,k_2)\, X^M(k_1,k_2)\,,
\]
where the deformed structure constant arises from
\[
\{e^{ik_1\cdot x},e^{ik_2\cdot x}\}^M = X^M(k_1,k_2) \,e^{i{(k_1+k_2)}\cdot x}\,,
\]
and is given by
\[
X^{M}(k_{1},k_{2})= \frac1{\alpha} \sinh\big(\alpha\, X(k_{1},k_{2})\big) \,.
\]
One can also define the object
\[
\label{eq:LkM}
{ L_k^M}:=\{e^{i\,k\cdot x},\cdot\}^M= \frac{e^{i\,k\cdot x}}{\alpha} \sinh\big(\alpha\, k_u\,\epsilon^+\!\cdot \partial\big)
\]
such that\footnote{To verify this formula, it is convenient to use the identity\\
$
\sinh(A\cdot\partial)\,e^{B\cdot x} = \sinh(A\cdot B)\,e^{B\cdot x}\cosh(A\cdot\partial)+\cosh(A\cdot B)\,e^{B\cdot x}\sinh(A\cdot\partial)
\,.$}
\[
\label{eq:LkMbracket}
[L^M_{k_1},L^M_{k_2}]= X^M(k_1,k_2)\,L^M_{k_1+k_2}\,.
\]

The physical content of this theory has a standard interpretation of describing SDG on a non-commutative background, where the non-commutativity is restricted to the null plane $(u,w)$.\footnote{This choice of non-commutativity plane is very relevant, leading to simplified equations in light-cone gauge, because the plane $(u,w)$ can be highlighted as a gauge choice in SDG, as we saw earlier.} That is the context in which it is studied e.g.~in \cite{Garcia-Compean:2003nix}, which belongs to a body of work that was motivated by the appearance of non-commutativity in M/string theory (see \cite{Douglas:2001ba,Szabo:2001kg} for reviews). In this standard interpretation, the effect of non-commutativity corresponds to the addition of Lorentz-breaking higher-derivative corrections to the spin-2 Lorentz-invariant theory of SDG.

Here, we want to propose an additional interpretation of the equations of Moyal-deformed self-dual gravity, that connects it to a theory of higher spins. The first clues are in the two Lie algebras at play in \eqref{eq:3ptvertMSDG}. The `left algebra' is generated by ${L_k}$ and carries spin 1, since ${L_k}\sim \epsilon^+\!\cdot \partial$. The `right algebra' is generated by ${L_k^M}$ and carries contributions from a tower of odd spins, each $\sim (\epsilon^+\!\cdot \partial)^\text{spin}$, all of which have fixed coefficient with respect to the spin 1 contribution.
 This suggests that the double copy between the two algebras can be associated to a constrained chiral higher-spin theory with even spins $\geq2$. By constrained, we mean that there are no more degrees of freedom than in SDG, since the higher spins are fixed by the graviton.\footnote{Notice that the generators ${ E_k^M}:= \frac{e^{i\,k\cdot x}}{2\alpha} \exp\big(\alpha\, k_u\,\epsilon^+\!\cdot \partial\big)$, which include all spins, lead to the same structure constants as \eqref{eq:LkM}. However, the former generators are singular as $\alpha\to0$. Nevertheless, this may indicate that the spin content of a higher-spin interpretation of Moyal-deformed SDG is not unique. Indeed, the chiral higher-spin theories \cite{Metsaev:1991mt,Metsaev:1991nb,Ponomarev:2016lrm} to be mentioned admit various versions/truncations, e.g. all spins versus excluding odd spins.}

Such an interpretation of the theory is consistent with the general picture of chiral higher-spin theories \cite{Metsaev:1991mt,Metsaev:1991nb,Ponomarev:2016lrm}, reviewed recently in \cite{Ponomarev:2022vjb}. The vertex \eqref{eq:3ptvertMSDG} is expanded for a small deformation as
\[
\label{eq:3ptvertMSDGexp}
V_\text{Moyal-SDG} = \frac1{\alpha^2}\, \sum_{\sigma\geq1}\, \frac{\big(\alpha X(k_1,k_2)\big)^{2\sigma}}{(2\sigma-1)!} \,.
\]
We can compare this to the chiral higher-spin (chs) vertices, written with our conventions as
\[
\label{eq:3ptvertchs}
V_\text{chs} = \frac1{\alpha^2}\, \frac{\big(\alpha X(k_1,k_2)\big)^{h_1+h_2+h_3}}{(h_1+h_2+h_3-1)!} \,,
\]
where the $h_i$ are the helicities of the incoming particles.\footnote{We take the higher-spin helicity fields to be redefined analogously to \eqref{eq:rescale}.}  The helicities are such that $h_1+h_2+h_3$ is positive and, due to Bose symmetry, even. This suggests that Moyal-deformed SDG may be obtained by identifying all the helicity fields with the graviton fields of the same helicity.
The vertex \eqref{eq:3ptvertMSDGexp} has contributions from all the vertices \eqref{eq:3ptvertchs} compatible with this. This interpretation is not without puzzling features. One important question is that Moyal-deformed SDG is not Lorentz invariant,\footnote{Recalling footnote \ref{footnote}, for a spin 2 field the external helicity factors are $\langle \eta i \rangle^{\mp 4}$. Given the form of the vertex \eqref{eq:3ptvertMSDGexp}, this makes it impossible to eliminate the dependence of the amplitude on the gauge reference spinor $|\eta\rangle$; it only works for the term $\sigma=1$ corresponding to SDG.} and yet its vertices are clearly associated to those of chiral higher-spin theories, which are Lorentz invariant (though not manifestly so in light-cone gauge, just like SDG). Another question is that while Moyal-deformed SDG has derivative corrections to unbounded order, chiral higher-spin theories obey a notion of locality \cite{Ponomarev:2016lrm}. We leave a more detailed investigation of the relation between Moyal deformations and chiral higher-spin theories for future work. Our discussion here is closely related to comments in \cite{Ponomarev:2017nrr}, where a version of \eqref{eq:LkMbracket} appears as equation (3.44), and in \cite{Ponomarev:2022atv}, where a non-local non-Lorentz-invariant `amplitude' obtained by summing over higher-spin 3-point amplitudes for all helicities is considered. The connection to chiral higher-spin theories highlights one feature: it is presumably impossible to construct a local parity-invariant extension of Moyal-deformed SDG (in contrast with undeformed SDG, whose parity-invariant version is general relativity). It is possible to construct interacting massless higher-spin theories in flat spacetime, but the examples exhibit limitations, e.g.~the higher-spin theory of \cite{Metsaev:1991mt,Metsaev:1991nb,Ponomarev:2016lrm} is chiral, and has vanishing tree-level scattering amplitudes.\footnote{See \cite{Skvortsov:2018jea,Skvortsov:2020wtf,Skvortsov:2020gpn} for some loop-level studies of chiral higher-spin theories. In the cases studied in these works, it is argued that the loop amplitudes vanish based on a regularisation of the infinite number of degrees of freedom.}

Chiral higher-spin theories have been mentioned recently in the celestial literature in connection with chiral OPEs \cite{Ren:2022sws}. It was checked in examples related to effective field theories that the chiral higher-spin vertices are consistent with OPE associativity. This work built on refs.~\cite{Himwich:2021dau,Mago:2021wje}, which considered deformations of celestial OPEs and of the algebra $w_{1+\infty}$ in the context of non-minimal couplings. It was found in \cite{Mago:2021wje} that a $W$-type deformation of this algebra arises when introducing non-minimal couplings, and that the OPE Jacobi identity in the cases studied dictates that interactions between gravitons and additional fields (e.g.~scalar) must be included. This is closely related to the Moyal-deformed SDG considered here, which, as we will see in the following, is related to a $W$ deformation of $w_{1+\infty}$. It is clear that there is more to explore, and it would be interesting to study the relation between these deformations, and to unveil the full picture of chiral higher-spin theories from a celestial perspective. It would also be interesting to investigate the relation between the results of \cite{Ren:2022sws} concerning vanishing amplitudes and the discussion in the next section.

Let us now investigate the counterpart of $w_{1+\infty}$ in the Moyal-deformed SDG. For that, we consider the soft expansion coefficients defined in \eqref{eq:softexp} as generators of the deformed Lie algebra. From \eqref{eq:Moyalexpl}, we obtain
\begin{align}
&\{ {\mathfrak e}_{a,b}\,,\, {\mathfrak e}_{c,d} \}^M \nonumber \\
& \qquad  = \sum_{s\geq0}\frac{\alpha^{2s}}{(2s+1)!}\sum _{j=0}^{2s+1}(-1)^{j}\binom{2s+1}{j}
\,
[a]_{2s+1-j} [b]_j [c]_j [d]_{2s+1-j}
\,{ {\mathfrak e}_{a+c-1-2s,b+d-1-2s}}\,.
\label{eq:W}
\end{align}
In this expression, \,$[a]_n=a!/(a-n)!$\, is the descending Pochhammer symbol, defined to vanish for $n>a$ so that the sum over $s$ is truncated. The analogues of \eqref{eq:OPEgrav} and \eqref{eq:OPEsoftgrav} are, therefore,
\[
{\mathcal O}_+(k_1)\,\, {\mathcal O}_+(k_2) \,\sim\, \frac{{ X^M(k_1,k_2)}}{z_1-z_2}\,\, {\mathcal O}_+(k_1+k_2)
\]
and
\begin{align}
& \varpi_{a,b}(z_1)\, \varpi_{c,d}(z_2) \,\sim \frac1{z_1-z_2}\; \times\nonumber \\
& \quad  \sum_{s\geq0}\frac{\alpha^{2s}}{(2s+1)!}\sum _{j=0}^{2s+1}(-1)^{j}\binom{2s+1}{j}
\,
[a]_{2s+1-j} [b]_j [c]_j [d]_{2s+1-j}
\;{ \varpi_{a+c-1-2s,b+d-1-2s}}(z_2)\,.
\end{align}
We obtain a deformation of the wedge subalgebra of $w_{1+\infty}$. To compare to the $W$ deformations described in \cite{Pope:1989ew,Pope:1989sr,Pope:1990kc} and reviewed in \cite{Pope:1991ig,Shen:1992dd}, we start by identifying the generators $V^i_m$ there to $\{\frac1{2}\,{\mathfrak e}_{i+1+m\,,\,i+1-m}\,,\,\cdot\,\}^M$, and we also drop the central terms, which are eliminated by the wedge condition, translated now into $|m|\leq i+1$. If we do this, we notice that the structure constants match those in \eqref{eq:W} if the numerical coefficients $\phi^{ij}_{2s}$ appearing in \cite{Pope:1990kc} take the value 1, which indeed corresponds to a particular member of the $W$ family of deformations.\footnote{We note that previous versions of our paper contained erroneous statements about $W$ algebras. Ref.~\cite{Bu:2022iak}, which appeared shortly after our paper, correctly described the relation of the deformation of $w_{1+\infty}$ to the $W$ algebra literature.}

There had been recent speculation that a $W$ deformation may appear in SDG at the quantum level, being a `quantisation' of $w_{1+\infty}$. Their counterparts, the Moyal bracket and the Poisson bracket, are also related by a quantisation. However, the 2D quantisation of the algebra is not the same as the quantisation of the 4D spacetime field theory and, therefore, not the same as the quantisation of the 2D celestial theory, at least not in an obvious sense. Indeed, we discussed in this section the appearance of a $W$ deformation in a purely classical deformation of SDG, given in light-cone gauge by higher-derivative corrections. These corrections appear to be unrelated to spacetime quantum effects. The results in \cite{Ball:2021tmb} highlight the point we are making: it was found that $w_{1+\infty}$ persists undeformed at loop level in the celestial OPE of the quantum undeformed SDG.\footnote{After the first arXiv version of this paper, ref.~\cite{Shyam:2022iwd} proposed that the Moyal deformation of self-dual gravity can also be seen as a $T\bar T$ deformation.}

\section{Chiral OPEs and integrability}
\label{sec:int}

In this section, we wish to highlight the role of the kinematic algebra \eqref{eq:LieX} in the classical integrability of the theories discussed here, from the perspective of scattering amplitudes. The point is that the kinematic algebra ensures the vanishing of the tree-level amplitudes, and we will see how this works. 

Let us consider a generic chiral celestial OPE of the form
\[
\label{eq:opeC}
{\mathcal O}_I(k_1)\, {\mathcal O}_J(k_2) \,\sim\,  \frac{ C_{IJ}{}^{K}}{z_1-z_2}\,\, {\mathcal O}_K(k_1+k_2) \,.
\]
For SDYM, we have $C_{IJ}{}^{K}=f^{a_1a_2a_3}$; for SGD, we have $C_{IJ}{}^{K}=X(k_1,k_2)$; and for Moyal-deformed SDG, we have $C_{IJ}{}^{K}=X^M(k_1,k_2)$. We are considering theories of massless particles, with kinematics described as in the beginning of section~\ref{sec:SD}. We can, therefore, write
\[
\label{eq:opeCs}
{\mathcal O}_I(k_1)\, {\mathcal O}_J(k_2) \,\sim\,  \frac{{ X(k_1,k_2)}\,{ C_{IJ}{}^{K}}}{s_{12}}\,\, {\mathcal O}_K(k_1+k_2) \,.
\]

Suppose that such a chiral system of OPEs fully determines the theory. This is the case for the three examples mentioned, which have only the vertex $(++-)$. Yang-Mills theory and gravity beyond the self-dual sector possess non-chiral OPEs, and therefore are not in the purely chiral class that we consider here.\footnote{Recall footnote \ref{footnote:chiral}. We will mention in the Discussion section that there are parity-asymmetric ways of constructing full Yang-Mills theory and gravity that look chiral, but nonetheless have crucial non-chiral elements.} The chiral class has a double copy structure where the left copy is associated to $X(k_1,k_2)$ and the right copy is associated to $C_{IJ}{}^{K}\,$.\footnote{In the standard terminology of the double copy, the left copy is SDYM. The double copy decomposition of SDYM itself is that the left copy is SDYM and the right copy is the bi-adjoint scalar theory, which has two colour-type Lie algebras. The double copy procedure `eats up' one copy of colour from each of the left and right copies, leading to SDYM.} In the examples of SDYM and (Moyal-)SDG, the OPE written as \eqref{eq:opeCs} is directly associated to the 3-point vertex in an off-shell formulation of the theories which has no higher-point vertices, but this identification is too restrictive in general.\footnote{For instance, $C_{IJ}{}^{K}$ could be associated to the full kinematic algebra, instead of the self-dual restriction $C_{IJ}{}^{K}=X(k_1,k_2)$ occurring in SDG. In that case, we are dealing with a sector of the 4D Einstein--dilaton--B-field gravity that, by the arguments in this section, has vanishing amplitudes and is, therefore, expected to be classically integrable.} We are assuming associativity of the OPE, at least at tree level as our argument applies to the classical theory. Associativity implies that the objects $C_{IJ}{}^{K}\,$ are constrained by the OPE Jacobi relations, which are identified in the simple examples of self-dual theories as BCJ-type Jacobi relations. More generally, at least naively, the latter appear to be a stronger condition due to the collinear kinematic restriction in the OPE. It would be important to clarify this point, but for now we will proceed in the assumption that the BCJ Jacobi relations can be satisfied in some representation of the `right copy theory' associated to $C_{IJ}{}^{K}$, which is a condition for the KLT formula \cite{Kawai:1985xq} below to hold.  

The idea now is that, for the cases of the chiral class discussed in the previous paragraph, the $n$-point tree amplitudes can be computed from the KLT formula associated to the double copy:
\[
{\mathcal A}_n = \sum_{\rho,\rho'\in S_{n-3}}
{ A}_{\text{left}} (1,\rho,n-1,n)
\;S_\text{KLT}[\rho|\rho'] \; { A}_{\text{right}}(1,\rho',n,n-1)\,.
\]
Here, the sums are over all permutations of the set of particles excluding $\{1,n-1,n\}$; $S_\text{KLT}$ is the momentum kernel, whose components are polynomials of order $n-3$ in the Mandelstam variables; ${ A}_{\text{left}}(\cdots)$ are the partial amplitudes with the prescribed ordering of the external legs, built with $X(k_1,k_2)$ as a vertex;  and ${\tilde A}_{\text{right}}(\cdots)$ are, analogously, the partial amplitudes built from the `right algebra' associated to $C_{IJ}{}^{K}$\,.

The details of the KLT formula are not very important here. What matters is that the partial amplitudes ${ A}_{\text{left}}(\cdots)$ built from the kinematic algebra $X(k_1,k_2)$ all vanish, leading to the vanishing of the tree amplitude ${\mathcal A}_n$. For instance, at 4 points, there are two diagrams contributing to  ${ A}_{\text{left}}(1,2,3,4)$, and it is easily checked using momentum conservation that this partial amplitude vanishes:
\[
\frac{X(k_1,k_2)\,X(k_3,k_4)}{s_{12}} + \frac{X(k_2,k_3)\,X(k_4,k_1)}{s_{23}} = 0\,.
\]
The proof that these partial tree amplitudes vanish at any multiplicity\footnote{With the caveat of footnote \ref{footnote3pts}.} is a basic result in the amplitudes literature, which can be obtained in various ways, e.g.~using Feynman diagrams with a smart gauge choice, using the Berends-Giele recursion, or using the supersymmetric Ward identities; see e.g.~ref.~\cite{Elvang:2015rqa}. This proof holds for generic kinematics, not for special (complexified) kinematic configurations where the amplitudes have support on $s_{ij}=0$ due to holomorphic collinearity ($\lambda_i$ all equal) as discussed in \cite{Witten:2003nn}; see e.g.~\cite{Ponomarev:2022atv} for a recent higher-spin example of this feature. 

The argument presented in this section, already made implicitly to some extent in \cite{Chacon:2020fmr}, is a scattering amplitudes version of how the Ward conjecture \cite{Ward:1985gz} works. The conjecture states that all classically integrable systems are some `reduction' of SDYM, which in our language is represented by the fact that the left copy in the double copy is the self-dual kinematic algebra \eqref{eq:LieX}.

\section{Discussion}
\label{sec:discussion}

In this paper, we drew a connection between the double copy, especially its formulation in terms of the BCJ colour-kinematics  duality, and the structure of the celestial chiral OPEs. The double copy decomposition of these OPEs is as follows: the `left copy' comes from the implicit area-preserving diffeomorphism algebra arising solely from the OPE chirality; the `right copy' comes from the structure constants of the OPE, which at least in the examples considered were identified with BCJ-type structure constants of a colour/kinematic algebra.

We showed that this connection between the double copy and the celestial OPEs provides an interpretation of the classical integrability of theories fully determined by chiral OPEs, namely a scattering amplitudes version of the Ward conjecture.

Our investigations were partly motivated by the celestial appearance of (the wedge subalgebra of) $w_{1+\infty}$ following refs.~\cite{Guevara:2021abz,Strominger:2021lvk}. We discussed how this algebra arises straightforwardly from the soft expansion of the self-dual kinematic algebra appearing in the `right copy' of the double copy decomposition of self-dual gravity. 
This point of view is a spacetime counterpart to old and recent work that describes self-dual gravity in twistor space \cite{Adamo:2021lrv}. Following the same approach, we described the deformation of $w_{1+\infty}$ associated to a Moyal deformation of self-dual gravity. It would be good to connect this result to those of \cite{Himwich:2021dau,Mago:2021wje,Ren:2022sws}, and more generally, to understand the space of consistent deformations and its relevance.

One important question is what role $w_{1+\infty}$ plays in the full theory, i.e.~beyond the self-dual sector. It may be worth it to put this question into the more general context of `chiral' and `non-chiral' approaches to scattering amplitudes in four spacetime dimensions. The following is by no means an exhaustive survey, but it may benefit some readers less familiar with these topics.
\begin{itemize}

\item Non-chiral (parity-symmetric) approaches: the chiral and anti-chiral sectors appear on equal footing. This is the case for the standard Yang-Mills theory and gravity Lagrangians, where we have both the $(+,+,-)$ and the $(-,-,+)$ vertices, complemented by higher-point vertices required by gauge invariance. The `distance' from the chiral/anti-chiral sectors may be measured by the number of vertices beyond that sector required by a given object, say an amplitude with certain helicities. This is related to the notion of MHV degree or (next-to)${}^k$-MHV, where MHV (maximal-helicity violating) amplitudes are the simplest, and at tree level should be thought of as the next-to-self-dual amplitudes sector (the first non-vanishing), where each diagram requires one non-$(+,+,-)$ vertex. For instance, the colour-kinematics duality for the MHV sector essentially inherits the self-dual kinematic algebra \cite{Monteiro:2011pc,Boels:2013bi}. Generically, the further an object is from both the chiral and anti-chiral sectors, the more complicated it is. This holds irrespective of the formulation, say irrespective of the use of vertices, as in the BCFW recursion relation, where the basic building blocks are the two types of 3-point amplitude \cite{Britto:2005fq,Arkani-Hamed:2012zlh}. An approach that provides a beautiful symmetric splitting between the chiral and anti-chiral sectors of amplitudes is the 4D ambitwistor string \cite{Geyer:2014fka}, but the explicit construction of the complete amplitude is still not straightforward, due to the sum over solutions to the polarised scattering equations. Polarised or not, the 4D scattering equations become harder to solve as the MHV and $\overline{\text{MHV}}$ degrees grow \cite{Roiban:2004yf,Cachazo:2012da,Cachazo:2013gna}. Fortunately, solving them explicitly is not necessary when studying certain properties of the amplitudes.

\item Chiral approaches: the theories are formulated in an `almost chiral' manner, where a systematic expansion away from the chiral sector is achieved by the introduction of non-chiral elements. A basic example is the CSW recursion \cite{Cachazo:2004kj} born out of twistor string theory \cite{Witten:2003nn}. Here, `MHV vertices' are added to the self-dual theory to provide non-vanishing tree amplitudes of any MHV degree desired \cite{Mansfield:2005yd}; the more of these vertices are used, the higher the MHV degree, and the more complicated is the result. Various twistor space approaches, naturally tuned to the self-dual sector, fall into this class; see \cite{Adamo:2021bej,Adamo:2021lrv,Costello:2022wso} for recent examples. For instance, in the case of \cite{Costello:2022wso}, the tree-level amplitudes are computed as a type of form factor, where the external particles are represented by elements of a celestial chiral OPE, and the operator in the form factor is analogous to the CSW MHV vertices. A very different instance of chirality arises in the ambitwistor string \cite{Mason:2013sva}, even in the 4D case mentioned above. This is the chirality of the worldsheet model (in contrast with conventional string theory). Here too, the worldsheet model is chiral except for the insertion of vertex operators which introduce localised non-chirality, and which are associated to the scattering equations. 

\end{itemize}

The preceding discussions in this paper suggest that the importance of the algebra $w_{1+\infty}$ in the chiral sector does not extend straightforwardly to the full celestial theory, because there is a price to pay as one deviates from this classically integrable sector. We saw that this algebra results from the soft expansion of the self-dual kinematic algebra, which explains the colour-kinematics duality in the self-dual sector but not in general. MHV amplitudes inherit much of the structure of the self-dual theory, and the self-dual kinematic algebra still explains the colour-kinematics duality in this next-to-self-dual sector. For these and other reasons, it is natural to investigate the role of $w_{1+\infty}$ starting from the MHV sector. However, the difficulty in finding a fully satisfactory description of the complete BCJ kinematic algebra (beyond MHV) is probably a good measure of the sophistication required of a parent algebra of $w_{1+\infty}$ that applies to the full theory. In fact, it is not inconceivable that the answer to both questions is the same in some form.

\section*{Acknowledgements}

It is a pleasure to thank Alfredo Guevara, Hongliang Jiang, Lionel Mason, Andrea Puhm and Atul Sharma for discussions, and Erick Chac\'on, Hugo Garc\'ia-Compe\'an, Andr\'es Luna, Chris White and Sam Wikeley for collaboration on related topics. This work was supported by a Royal Society University Research Fellowship.

\bibliography{refs}
\bibliographystyle{JHEP}

\end{document}